\documentclass[12pt]{article}

\usepackage{graphicx}

\usepackage[cmex10]{amsmath}
\usepackage{amsfonts}
\usepackage{amsthm}

\usepackage{url}

\usepackage{fixltx2e}

\theoremstyle{definition} \newtheorem{theorem}{Theorem}[section]
\theoremstyle{definition} \newtheorem{definition}[theorem]{Definition}
\theoremstyle{definition} 
\theoremstyle{definition} 
\theoremstyle{definition} 
\theoremstyle{definition} 
\theoremstyle{definition} 
\theoremstyle{definition} 
\theoremstyle{definition} \newtheorem{remark}[theorem]{Remark}
\theoremstyle{definition} 
\begin{document}

\title{The Potluck Problem}
\date{}
\maketitle

\begin{center}
\author{\begin{tabular}[t]{c@{\extracolsep{2em}}c}
    Prabodh Kumar Enumula & Shrisha Rao \\
    {\tt prabodh.kumar@in.ibm.com} & {\tt srao@iiitb.ac.in} \\
    IBM India Pvt. Ltd. & IIIT - Bangalore \\
\end{tabular}}
\end{center}

\begin{abstract}

  This paper proposes the Potluck Problem as a model for the behavior
  of independent producers and consumers under standard economic
  assumptions, as a problem of resource allocation in a multi-agent
  system in which there is no explicit communication among the agents.

\end{abstract}

\noindent{\bf Keywords:} weighted majority algorithm, Santa Fe Bar Problem, 
demand-supply parity, rational learning, predictors

\noindent{\bf DOI:} 10.1016/j.econlet.2009.12.011

\section{Introduction}

In the study of bounded rationality and inductive reasoning, Brian
Arthur introduced the \emph{Santa Fe Bar problem}
(SFBP)~\cite{Arthur1994}. The SFBP deals with the allocation of
constrained resource to non-cooperating multiple agents.  SFBP
extensions have been studied in resource allocation games by different
authors.  For instance, Schaerf et al.~\cite{Schaerf1995} studied
multi-agent learning in context of adaptive load balancing, while
Galstyan et al.~\cite{galstyan03resource} studied resource allocation
games with changing resource capacities.  In the model considered by
Galstyan et al., there is communication among the agents before
choosing their strategy.  But here in the model we are proposing,
there is no explicit communication among the agents.  In this paper we
generalize the SFBP by introducing multiple producers and also by
considering varying demands for the resources.  We propose the
\emph{Potluck Problem} (PP) to model resource allocation and
utilization in a multiple-producer, multiple-consumer environment.
The model depicted by the PP is applicable in many real-world
situations, e.g., an electrical power grid with many individual power
supply units and consumers of power.  An economy in which there are
multiple agents who predict the global behavior and take local
decisions can be modeled.  In a standard multi-player economic
environment, price allocation in the presence of varying demands for
the resource is a situation which resembles PP.  Many distributed
systems such as water management systems, Internet service providers
where service on demand is required, etc., can be modeled by the PP.

This problem is modeled by observing multiple instances of potluck
dinner by a set of people.  A potluck is a gathering of people where
each person is expected to bring a dish of food to be shared among the
group.  The multiple persons (agents) will decide individually what
quantity of food to contribute to the dinner, without any prior
coordination among themselves.  Such instances of dinner are repeated.
The problem in deciding how much to contribute to the dinner is
because of the varying demand for the food in every instance of the
dinner.  (For simplicity, we consider all food as consisting of just
one dish.)

Section~\ref{problem} describes the Potluck Problem in more detail and
also gives an explanation to how it is considered as generalizing the
SFBP.  In Section~\ref{impossibility}, it is shown that with rational
learning, equilibrium state cannot be achieved in the PP.  Then a
weighted majority~\cite{littlestone89weighted} learning algorithm is
applied to achieve near-equilibrium behavior.
Section~\ref{conclusions} discusses results of simulation of the PP.  
In the last section, possible extensions of this work are discussed.

\section{Problem Description} \label{problem}

Consider a scenario where $N$ (e.g., 100) people (or agents) have a
potluck dinner every week in a city.  Each of them decides
individually how much food to contribute to the dinner.  The agents do
not communicate among themselves except that every agent knows about
the total demand and supply at the dinners in previous weeks.  The
dinner is enjoyable if there is no \emph{starvation} or \emph{excess}
of food.  Starvation means that the amount of food brought to the
dinner is not sufficient to serve the people in the dinner, while
excess means that some food goes waste for being more than sufficient
for the dinner.  The demand for food by each agent varies according to
variable individual appetite each week.  An agent decides on how much
to contribute depending on the prediction it makes for the demand and
supply in that week.

Specifically, as described we have a \emph{supply-side} problem, but a
corresponding \emph{demand-side} problem can also be formulated where
the available supply of a resource varies and demand is to be adjusted
accordingly.  The demand-side problem has applications in electrical
power technologies, where it is called \emph{demand
  response}~\cite{SpeesLave2007a}, and elsewhere.  The model and
results are very similar, however, so we do not describe this in
detail.

In a game-theoretic fashion, the Potluck Problem can be described as a
repeated, non-cooperative game.  Say there are $N$ agents who are
players in the game.  Consider one instance of the game (one
``dinner''), $t$.  For a player \(i \leq N\), the strategy set is \(0
\leq Q_i \leq Max_{i}\), where $Q_i$ is the quantity of food carried
to a dinner by agent $i$, and $Max_{i}$ corresponds to going to dinner
with the maximum supply of food that that agent $i$ can take.  Let
$M_i$ denote the set of probability distributions over $Q_i$ which
defines the mixed strategy for agent $i$.  Now say \(s_{i,t} \in M_i\)
indicate the mixed strategy of the player $i$ at instance $t$ of
dinner, then the total source of food to the dinner will be \(S_{t} =
\sum_{i=1}^n s_{i,t}\).  The agent decides on the mixed strategy
$s_{i,t}$ by predicting the total demand for the dinner which is
denoted by $P_{i,t}$.

In every dinner all/some of the agents also act as consumers by
consuming the food brought to the dinner.  The agents' demand for the
food also varies over different dinners.  The demand for food by an
agent $i$, at an instance $t$ is given by $d_{i,t}$. So the total
demand for food in the dinner is \(D_{t} = \sum_{i=1}^n d_{i,t}\).
Ideally, the dinner is enjoyable if \(S_{t} = D_{t}\).  This state of
the game, where the supply and demand exactly match, is the
equilibrium state in the Potluck Problem.  If \(S_{t} < D_{t}\), then
there is \emph{starvation} at the dinner, and if \(S_{t} > D_{t}\)
there is an \emph{excess}.  The Potluck Problem is a repeated game of
such instances.

\begin{remark}
  The Potluck Problem is a generalization of the SFBP.
\end{remark}

Consider a case of the Potluck Problem in which the demand for the
resource is fixed over all iterations of the dinner, i.e., \(\forall
t, D_t = d\) such that \(d < N\).  Also assume the strategy set for
all agents is constrained to take discrete values, i.e., \(Q_i =
\{0,1\}, \, i \leq N\). Now at an instance $t$ of the dinner, if \(S_t
< D_t\), then all players will increase their supply by choosing
\(s_{i,t} = 1\).  This will result in making the total supply to the
dinner more than the demand, i.e., \(\sum_{i=1}^n s_{i,t} = N\), which
will cause an excess (\(S_{t} > d\)) .  Then all players will choose a
strategy \(s_{i,t} = 0\) for the next dinner, causing starvation.
This kind of oscillatory behavior is similar to that in the Santa Fe
Bar Problem.

To observe the similarity with the SFBP, consider \(d = 60\) for all
instances of the game. For each player the strategy set \(Q_i =
\{0,1\}\) corresponds to choosing not to attend the bar and choosing
to attend the bar respectively.  An under-crowded bar is one in which
there is a resource (place in the bar) which is going waste.  This is
similar to excess in the PP.  An overcrowded bar likewise resembles
starvation in the PP.  Thus the Potluck Problem is a generalization of
the SFBP.

\section{Impossibility of Rational Learning} \label{impossibility}

The oscillatory behavior that is known to arise the PP shows the need
to have some predictive mechanism by which agents can foresee the
demand for the coming dinner and then decide upon on the supply they
want to bring to achieve the equilibrium in the PP.

\subsection{Best-Reply Dynamics and $\epsilon$-Predictive Learning}

This section formalizes the argument pertaining to oscillatory
behavior observed in the PP.  First we note the standard best-reply
dynamics and then explain the same in the context of the PP.

The best reply dynamics is often termed as the Cournot adjustment
model or Cournot learning after Augustin Cournot who first proposed it
in the context of a duopoly model~\cite{Cournot1838}.  The best reply
dynamics can be seen as a game in which each self-interested agent
assumes that every other similar agent uses the same strategy in later
periods that is similar to one most recently used.  In similar vein, a
learning algorithm used by an agent is \emph{predictive} if it
correctly matches up with the situations created by the agents.

In the present context, we can say the following.

\begin{definition}

\begin{itemize}

\item[(i)] An agent $i$ is said to be employ the best-reply dynamics
  in the PP, iff for all $t$ , the player assumes that $P_{i,t} =
  D_{t-1}$ and decides on the supply $s_{i,t}$ the player wants to
  contribute.

\item[(ii)] A learning algorithm is said to be
  $\epsilon$-\emph{predictive} in the PP, iff it generates a sequence
  of beliefs $P_{i,t}$ for a player $i$, such that \(| P_{i,t} - D_{t}
  | \leq \epsilon\), for some $\epsilon \geq 0$.

\end{itemize}

\end{definition}
 
If we say agent $i$ is able to predict using a learning algorithm,
then the difference between its prediction about the next week's
demand and the actual demand for that particular week should be zero,
or at least less than some $\epsilon$. An agent $i$ is said to be
\emph{rational}, iff it plays only best-replies to its
beliefs~\cite{greenwald98santa}. When playing rational, the agent
assumes that the demand for the coming week is going to be same as the
previous week, i.e., \(P_{i,t} = D_{t-1}\).

\begin{remark}

  In the PP, if all the agents are rational and $\epsilon$-predictive,
  then \(\forall t, \, | D_{t} - D_{t+1} | \le \epsilon\), i.e., the
  variation in total demand between successive dinners should at most
  be $\epsilon$.

\end{remark}

However, given that the variation in total demand need not be bounded
by $\epsilon$, it thus follows that rational agents will not be
$\epsilon$-predictive.  As a consequence, we have the following.

\begin{remark}
  In the PP, it is impossible to achieve equilibrium if agents employ
  rational learning.
\end{remark}

This in turn illustrates the need for agents to learn using better
principles than best-reply dynamics. Such learning is called
non-rational learning.

\subsection{An Algorithm for Non-Rational Learning}

As discussed in the previous section, a perfect rational approach may
lead to oscillations in systems which resemble the behavior predicted
in the Potluck Problem. So the agents in the game need a learning
which is not perfectly rational.

\begin{definition}
  A \emph{predictor} makes use of the previous dinner's data available
  and makes a prediction for the demand of the dinner for the coming
  week. It is a function which uses $S_{t-1} $,$S_{t-2}$ ,...$S_{t-x}$
  (past $x$ week's data) and $D_{t-1} $,$D_{t-2}$,...$D_{t-x}$ and
  predicts the demand for the coming week.
\end{definition}

Various predictors that can be considered by an agent in the PP
include:

\begin{itemize}
\item Average demand over the last $j$ (e.g., 10) instances of dinner.
\item Randomly choose the demand of one of weeks from the last $j$
  (e.g., 10) instances.
\item The rational predictor (presume that consumption the next time
  will be the same as the last time).
\item An oracle which gives consumption.
\item A time-varying function.
\end{itemize}

The initial choice of predictors depends on the kind of system we are trying 
to model.  In some, a time-varying function of demand should be incorporated 
as a predictor, e.g., in forecasting the demand for electrical power (which is 
typically higher during daytime than at night).  In others, an oracle which 
forecasts demand (based on data from outside the system not known to agents) 
may be appropriate.

Out of the various predictors available, each agent randomly chooses $k$ 
predictors.  Then each agent has $k$ predictions about the demand at the 
coming dinner, each of which is denoted by $O_{i,p,t}$, representing the 
prediction made by agent $i$'s $p^{th}$ predictor for dinner $t$.  The agent 
$i$ decides the supply $s_{i,t}$ to be taken to the dinner $t$ based on the 
forecasts of those $k$ predictors, using a weighted majority 
approach~\cite{littlestone89weighted}.  Each agent $i$ maintains a weight 
$W_{i,p,t}$ for each predictor $p$ at time $t$, and updates it after each 
iteration of the game, with the weight of accurate predictors increasing and 
that of inaccurate ones decreasing. The initial weights of all predictors may 
be equal or some random non-zero values.

The iterative update and learning algorithm used by the players can be
summarized in the following steps.

For an agent $i$, during dinner $t$, do the following.

\begin{itemize}
\item Predict demand using all the predictors, i.e., find
  \(O_{i,p,t}, p = 1,2, \ldots, k\).
\item Predict the demand $P_{i,t}$ by using all the predictions, using
  a weighted majority algorithm.
\item Decide on the supply $s_{i,t}$ to be brought to the dinner.
\item Update weights of all the predictors based on the actual
  demand and supply at dinner $t$.
\end{itemize}

The prediction for a particular instance of dinner is calculated by
taking the \emph{weighted majority}~\cite{littlestone89weighted} of
the predictions made by the all predictors of the agent, i.e.,
\[P_{i,t} = \frac{\sum_{p=1}^k(W_{i,p,t} \times
  O_{i,p,t})}{\sum_{p=1}^kW_{i,p,t}}.\] After every instance of
dinner, an agent updates the weights of all its predictors based on
how close they were to predicting the actual demand.

The update equation of the weight of a predictor $p$ at an instance
$t$ is \(W_{i,p,t+1}=W_{i,p,t} \times F\), where \(F=\beta^\upsilon\), with
$\beta$ being a parameter chosen such that \(0 < \beta < 1 \).  If
\(\frac{O_{i,p,t}}{D_{t}} > 1\), then $\upsilon$ is set to
\(\frac{O_{i,p,t}}{D_{t}}\), and if \(\frac{O_{i,p,t}}{D_{t}} \leq
1\), then \(\upsilon \leftarrow \frac{D_{t}} {O_{i,p,t}}\).

After updating all the weights, they are normalized to between 0 and 1
using the equation 
\[W_{i,p,t+1} = \frac{W_{i,p,t+1}}{\sum_{p=1}^kW_{i,p,t+1}}, \ p =
1,2, \ldots, k.\]

\begin{figure}[h!tbp]
\centering
\includegraphics[scale=0.5]{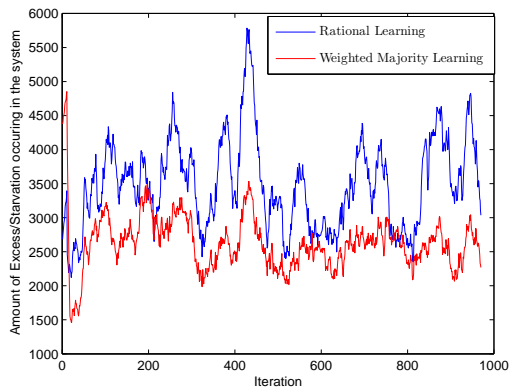}
\caption{Resource excess/starvation occurring in the PP}
\label{figure1}
\end{figure}

\subsection{Simulation Results}

The Potluck Problem has been simulated with 100 non-identical agents
with production capacities in the range of 500 to 1000 discrete units,
and consumptions in the range of 0 to 1000, with 1000 instances of
dinner. The mean consumption of all agents together came to 48,200.
The mean production under rational learning was 47580, and under
weighted majority learning was 48284.  The weighted majority approach
with the five simple predictors listed previously outperformed the
rational approach about 99.5\% of the time, and resulted in a level of
starvation/excess that was 22.6\% better than in the rational approach
on average, and about 41.5\% in the best case.  (More intricate
predictors are seen to yield even better results.)  The results are
depicted graphically in Figure~\ref{figure1}.

\section{Conclusion and Future Work} \label{conclusions}

This paper proposed and analyzed the nature of the Potluck Problem.
It is observed that a weighted majority learning approach results in
better parity between demand and supply, compared to rational
learning.  Though in the present work, we have modeled the problem as
only one resource being allocated/consumed, the problem can be easily
extended by considering multiple resources, which are needed in
specific, possibly unforeseen, proportions for better utility.  In the
current analysis, each agent chooses predictors from a pool of
predictors available.  This problem can be studied also by each agent
choosing a different set of predictors from others, which closely
resembles social behavior, and different sets of predictors can be
compared and evaluated.  Various predictor behaviors and performance
can also be studied for specific patterns of demand, which can help in
applying this model in a realistic scenarios.

\end{document}